\documentstyle[12pt,aps]{revtex}
\begin{document}
\title{
A note on the Georgi Vector Limit at finite temperature/density\bigskip}
\author{ Boris Krippa \thanks{on leave from the
Institute for Nuclear Research of
 the Russian Academy of Sciences, Moscow Region 117312,
Russia.}}
\address{Department of Physics, UMIST,
PO Box 88, Manchester, M60 1QD, UK and University of
Manchester, Manchester M13 9PL, UK  \\}
\maketitle
\bigskip
\vspace{-5.5cm}
\vspace{5.5cm}
\begin{abstract}
The mechanism of chiral symmetry restoration in the context
of the  Georgi Vector Limit (GVL) at finite
 temperature/density is discussed.
 It is suggested that, whereas the system may be driven to the GVL,
 the fact of reaching the point of restoration  does not
 necessarily mean the complete realization
of the GVL  since the other mechanisms of restoration
may contribute at lower temperatures/densities.

\end{abstract}
\newpage
 The statement that
 chiral symmetry is spontaneously broken in QCD is
confirmed by the experimental data and there can be no doubt about it.
The QCD ground state does not completely share the symmetries of the
QCD Lagrangian so that Goldstone bosons and a number of nonzero order
 parameters appear. However, the experience borrowed from the condensed
matter physics tells us that under some circumstances the
continuous  broken symmetry (in our case chiral symmetry) can be restored.
The restoration may occur either in the academic limit
of large $N_f$ \cite{Ba} or in the hadronic system at
finite (and high enough) temperature/density. In the limit
of the restored chiral symmetry the corresponding order parameters must
vanish. One of the possible order parameters which we will focus on is
the difference between the correlators of the chiral partners.
Namely we will consider the difference
$\Pi_V - \Pi_A$ between the vector and axial correlators.
There have been proposed several different scenarios of how
a vanishing $\Pi_V - \Pi_A$ difference might occur
at finite temperature/density \cite{Sh},\cite{Kr1}. In  the standard  picture
one of the possibilities is that
the masses of the chiral partners $\rho$ and $a_1$ become degenerate
at the critical point thus providing the vanishing order parameter.
A somewhat different variant of the standard pattern assumes
that both the vector and axial correlators exhibits peaks of the same
 strength at both the mass of the $\rho$ and $a_1$ so that their difference
should also vanish at the point of restoration. In the other scenario
suggested in \cite{Ge} and called the Georgi Vector Limit (GVL) the $\rho$
meson is treated as the gauge boson of the Hidden Local Symmetry (HLS)
which gets its mass through the Higgs mechanism. When the system
is driven closer to the critical point the $\rho$ meson gets lighter
and eventually massless. Accordingly, the $\rho$ meson (its longitudinal
component)  becomes the chiral partner of pion. So the  longitudinal
component of the $\rho$ meson can be treated as an effective
massless scalar. Thus, the GVL implies the existence of the relations

\begin{equation}
<0|V_{\mu}|\rho_L(p)> = i p_\mu f_s,
\end{equation}
and $f_s = f_\pi$, where $f_\pi = 93$ MeV  is the the standard weak decay constant.
It is worth mentioning that these couplings are not neceserily vanishing
in the unbroken $SU(3)\times SU(3)$ phase. This scenario is called
Vector Realization (VR) of the symmetry restoration. The other closely related
mechanism called ``Vector Manifestation'' (VM) was suggested in \cite{Ha}.
In the VM the relation $f_s = f_\pi = 0$ holds at the point
of restoration. The VM also requires GVL
to be realized in the chirally restored phase. The major difference
between the VR and VM mechanisms is that in the VR the Goldstone modes
exist, whereas chiral symmetry is not broken. In the VM mechanism
the Goldstone bosons disappear since $f_\pi = 0$.

One notes that in the standard scenario of chiral symmetry restoration
in some academic limit in vacuum the
 mass of the $\rho$ meson can  either be massless or massive but should
always be degenerated with the mass of the chiral partner. This is what
 happens in the limit of large  $N_f$. However, the situation becomes
more complicated at finite temperature/density. In this case the general
expression for the hadron correlator, satisfying the chiral
symmetry requirements, can be written as follows

\begin{equation}
\Pi_{\mu\nu}=\sum_{a,b}\int {d^3\hbox{\bf k}\over 2\omega_k} n_{\pi}(T/\rho,k)
\int d^4x e^{ip\cdot x}\langle A\,\pi^a(\hbox{\bf k})|T\{J(x)
J(0)\}|A\,\pi^b(\hbox{\bf k}')\rangle.
\label{piinout}
\end{equation}
Here $n_{\pi}(T/\rho,k)$ is the in-medium pion distribution
 which is the thermal
 phase factor in the finite temperature case and the average
number of pions in the case of nuclear matter. We denoted $J(x)$
 the corresponding
interpolating current. The states labeled as $|A>$ can either represent
the Fermi-gas of noninteracting nucleons or the hadron system in the
 thermal bath
with pionic contributions subtracted. At low temperature
the hadron medium is primarily formed by the thermal pions so the state
$|A>$ is just the QCD vacuum so that $|A\pi> \rightarrow |\pi>$.
 One notes that in the
case of finite baryon density pions are virtual so that all
 time directions must be taken into account.
Using the soft pion theorem and current algebra
commutation relations  one can get the chiral mixing between
axial and vector correlators

\begin{equation}
\Pi_V=\overcirc\Pi_V+\xi(\overcirc\Pi_A-\overcirc\Pi_V)
\label{PiV}
\end{equation}
and similar expression for the axial vector correlator $\Pi_A$
\begin{equation}
\Pi_A=\overcirc\Pi_A+\xi(\overcirc\Pi_V-\overcirc\Pi_A)
\end{equation}
Here  we defined (for the finite baryon density case)
\begin{equation}
 \xi (\rho)={4\rho\overline{\sigma}_{\pi{\scriptscriptstyle N}}\over 3f_\pi^2 m_\pi^2}.
\end{equation}
Here $\overline{\sigma}_{\pi{\scriptscriptstyle N}}$ is the pionic piece of the pion-nucleon
sigma term and $\rho$ is nuclear density.
The corresponding thermal factor is

\begin{equation}
 \xi (T)={T^2 \over 6 f_\pi^2}
\end{equation}
The mixing of the correlators is due to the in-medium pions interacting directly
with the vector/axial interpolating currents.
 We denoted $\overcirc\Pi_V$($\overcirc\Pi_A$)
the correlator of the vector (axial) currents, calculated without taking into account the
soft pion effects, leading to the chiral mixing. The other finite temperature/density effects,
like in-medium changes of masses or widths not related to the chiral mixing effect,
 may, in principle, be included
in the ``pionless'' correlators  $\overcirc\Pi_V$($\overcirc\Pi_A$).
The finite temperature mixing relation was derived in \cite{El} and the finite baryon density
case was considered in \cite{Kr}.
 Restoration of
chiral symmetry corresponds to the complete mixing, which means the equality
of the axial and vector correlators so that $\Pi_V - \Pi_A \rightarrow$ 0.
One notes in passing that Wigner realization of the chiral symmetry does
 not require the  pion
 and its chiral partner to be massless at the restoration point. The complete
 mixing and thus chiral symmetry
 restoration occurs at $T \simeq 0.16$ GeV and $\rho \simeq 3\rho_0$ where
$\rho_0 = 0.17 fm^3$ is the nuclear matter density.
 It is important to mention that, although
 formally the expressions above are valid at not very high temperature/density
the critical points as given by the mixing relations
 are surprisingly  close to those following from more sophisticated
models and lattice simulations. It means that the mixing relations reflect the general
model independent chiral structure of the in-medium correlators and approximately
 hold at densities/temperatures close to the critical. It could  qualitatively be understood
 as follows. The critical temperature $T_c \simeq 0.16$ GeV is still significantly lower
than the chiral scale $\Lambda \simeq 0.8 GeV$ so that the
chiral symmetry based arguments concerning the general  structure
of the in-medium correlators
may still be qualitatively applicable even at the temperatures not far from the critical.
Similar arguments hold for the finite density case.
In a sense, the
mixing relations can be interpreted as the leading order terms
in the chiral expansion of the pertinent correlators
where all nongoldstone degrees of freedom are collected in the ``pionless'' correlators
$\overcirc \Pi_V (\overcirc \Pi_A)$.

Considering the mixing relations in the context of the GVL one first notes that the GVL
can naturally be accommodated in the chiral  mixing relations. Indeed, the longitudinal part
of the axial correlator includes the contribution from the pion pole. We assume
that the in-medium pion remains  massless and focus basically on the finite
temperature case. The generalization to the finite density system is straightforward.
At low enough temperature the system is just lukewarm pion gas and
$\overcirc\Pi_V$($\overcirc\Pi_A$) are the vacuum vector(axial) correlators.
The axial correlator can be decomposed in the following way

\begin{equation}
\overcirc \Pi_{A,\mu\nu}  = - \left[ g_{\mu\nu} -
 {q_\mu q_\nu\over q^2}\right] \overcirc \Pi_1(Q^2)
- {q_\mu q_\nu\over q^2} \left[\overcirc \Pi_1(Q^2) + Q^2\overcirc \Pi_2(Q^2)\right]
\end{equation}
The longitudinal part of the axial correlator contains the contribution from
the pion pole
\begin{equation}
\overcirc \Pi_{A}^{L} \simeq {f_{\pi}^2\over Q^2}
\end{equation}
We assume that the mixing relation approximately holds even at
temperatures close to the critical one.  The vacuum pion decay
constant $f_{\pi}^2$ should, in principle, be  replaced  with
its finite temperature analog $f_{\pi}^2(T)$.

If the system is driven toward
the GVL than
\begin{equation}
\overcirc \Pi_V \simeq {f_{\rho}^2(T)\over Q^2}
\end{equation}
Where we assumed that the decay constant $f_\rho$ also acquires
some temperature dependence $f_{\rho} \rightarrow f_{\rho}^2(T)$.
At some critical temperature $f_{\pi}^2(T_c)= f_{\rho}^2(T_c)$ so that GVL is reached.
We note however that there is a competing possibility of reaching the
 chiral symmetry restoration point which is related to the correlator mixing.
Taking the difference $\Pi_V - \Pi_A$     one obtains

\begin{equation}
\Pi_V - \Pi_A \simeq {1 \over Q^2} \left[f_{\rho}^2(T) - f_{\pi}^2(T)
\right](1 - 2\xi (T)),
\end{equation}
so that the point $\xi (T)= 1/2$ could, in principle, be reached at the temperature
lower than that needed for reaching GVL. The physical consequence of this fact
is that, even if the system shows a tendency towards the GVL
 with increasing temperature,
the critical temperature needed for chiral symmetry
restoration does not neceserily coincides with that
at which the GVL is achieved and VM
$(f_{\rho}^2(T) = f_{\pi}^2(T) = 0)$ is implemented. Therefore, in the chirally
restored phase one still may have massless pions and light (but massive)
vector mesons. In a sense, the chiral mixing and VM are just
two competing scenarios, equally participating in the process of symmetry
 restoration. The other possible mode is the already  mentioned
VR mechanism.
Figuring out which mechanism will take over  eventually seems to be a quite
difficult problem which
may depend on the model used. Neither one can a'priori be excluded.
We stress that the above arguments are admittedly suggestive
and aim only at indicating the possibility that in the GVL scenario at finite
temperature/density the fact of reaching the GVL point implies the chiral
symmetry restoration whereas the opposite may not be true. One notes that
at $ T \rightarrow 0$ chiral symmetry restoration implies the mandatory realization of the GVL.

It is worth mentioning that the existence of a few possible scenarios within one general
pattern of restoration is also the case when the
standard mechanism (mixing of $\rho$ and its chiral partner $a_1$) is assumed.
At the point of restoration (the point of complete
mixing) the masses of the chiral partners may either
be degenerate,  different or become comparable with its widths.
Any of these possibilities is possible. One needs to
develop a sophisticated models
to understand which way is prefered by nature. Chiral symmetry alone
cannot provide the answer.
Similar thing may take place in the GVL. The symmetry can be restored
either in the VM or VR mode. It is possible that VR
 $(f_{\rho}^2(T) = f_{\pi}^2(T) \neq 0)$ is reached at the temperature
lower than that needed for the complete chiral mixing to occur.
Again, only a sophisticated enough model where the correlator mixing
and the possibilities of VM and/or VR modes are incorporated can provide
a reliable understanding  of the chiral symmetry restoration pattern in GVL.
The naive, leading order estimates, based on the chiral perturbation
 theory \cite{Le} give
\begin{equation}
 f_{\pi}^2(T)=  f_{\pi}^2(1 - {T^2\over 12f_{\pi}^2} + O(T^4))
\end{equation}
indicate that the
difference between axial and vector correlators approaches zero at
approximately the same temperatures as
 $f_{\pi}(T)$ does. The higher order $O(T^4)$ corrections
do not change this conclusion in any qualitative way.
These corrections were calculated in Ref. \cite {El1}. The part of them,
which is relevant in the context of this paper, amounts to the replacement
$\xi (T) \rightarrow \xi (T) - \xi^{2}(T)/2$ in the mixing relations.
It  gives somewhat higher ($T_c \sim 200$) MeV  but still reasonable critical temperature.
The  $f_{\pi}^2(T)$ shows the same tendency when the $O(T^4)$ corrections
are included. This behavior should, of course, be expected since both
the difference $\Pi_V - \Pi_A$ and decay constant $f_{\pi}(T)$ are the
order parameters. However, to understand the mechanism of
chiral symmetry restoration in the GVL one should also know the temperature
dependence of the  $f_{\rho}^2(T)$.
 In Ref. \cite {Ha} this constant
was parameterized as    $f_{\rho} = a f_{\pi}$. The successful
hadron phenomenology can be reproduced if the choice $a = 2$
is made. The GVL
 requires  $a(T_{GVL}) = 1$ at some critical temperature $T_{GVL}$.
 The temperature dependence
of the parameter $a$ was studied in \cite {Ha1} in the framework of the
HLS model. It was shown in \cite {Ha1} that
up to  $T \simeq 250 MeV$ this parameter exhibits very weak dependence
on temperature and almost coincides with the value $a = 2$
extracted from the vacuum phenomenology so at the temperatures
corresponding to the complete chiral mixing
the situation is still quite far from the GVL. Our independent
one-loop estimates resulted in the practically the same conclusion.
 We note, however,
that the one-loop calculations may not be
 completely sufficient at the temperatures close to the critical.
The other important aspect of the above discussion is the concept
of vector dominance (VD) at finite temperature/density. As demonstrated
in \cite {Ha1} VD is simply a consequence of a specific parameter
choice $a = 2$ in the HLS model. Therefore, if GVL mechanism
is indeed realized it implies that VD is badly violated at the point
 of restoration.

One possible way of studying the mechanism of restoration in the GVL is to use
lattice methods at finite temperature. Since the pion effects
can be included explicitly via chiral mixing and since pions remains almost
massless even at the critical point, so that soft pion approximation may
still be qualitatively correct, the temperature dependence  of the
 axial and vector correlators (with pion effects subtracted) could be
 calculated in the quenched or partially quenched approximation.
However, the lattice calculations are not really possible at finite
density so in this latter case some analytical models are required.

One needs to mention again that the arguments  given in this letter are
at best qualitative. A number of effects was not included. Besides the
 previously
mentioned higher order terms ($O(T^4)$ etc) the effects of the heavier mesons
 should  be included at
temperatures near critical. However, in our opinion it seems unlikely
that these corrections can qualitatively change the results
for at least $T \simeq 100$ MeV. The appearance of baryons at some finite temperature
may further complicate the situation. Baryons can roughly be considered as
``drops'' of chirally restored phase providing thus the other competing mechanism
of chiral symmetry restoration \cite{Ko}. However, at $T \simeq 100 MeV$
the baryonic gas is still quite dilute so the influence of baryonic degrees of
freedom on the mechanism of restoration seems to be  rather moderate. The
 constraints on the baryonic properties in the case of finite nuclear density,
following from chiral symmetry, were considered in \cite{Kr2}.

Let's summarize the main points made in this letter. As was noticed
in Ref. \cite {Ha} in the limit of large $N_f$ restoration of chiral symmetry
does imply the complete realization of the GVL and VM mechanism.
In the VM mechansim the massless $\rho$ meson (longitudinal part) becomes
the chiral partner of pion. So in this limit
 restoration of chiral symmetry and reaching GVL mean the same thing.
Such an equivalence between GVL and chiral symmetry restoration may not be the case
at finite temperature/density. There are some additional ``medium'' mechanisms of
chiral symmetry restoration which should be taken into account.
 At finite
temperature/density the hadronic system may also be driven toward the GVL so that
 vector mesons  become light. At some critical temperature the GVL is
reached and vector mesons turn out massless and degenerated with pions.
At this temperature  chiral symmetry is no longer broken.
So reaching of the GVL does mean the restoration of chiral symmetry.
The converse is not true at finite temperature/density. Even
if the system shows the tendency toward the GVL, the restoration of
chiral symmetry does not  mean that the GVL is reached.
The correlators of chiral partners get mixed in medium
and may, in principle, reach the point of complete mixing (where their
difference is zero) at the temperatures(or densities) lower than that needed for GVL
to be realized. Our crude estamates show that indeed the symmetry is restored at
temperatures/densities lower than that needed for GVL to be reached.
Therefore, unlike the large $N_f$ case,
the restoration  of chiral symmetry and reaching GVL may occur
at $different$  temperatures/densities.
 Moreover, the restored symmetry in the  GVL at finite temperature/density can
  manifest itself in the both  VM and VR
 modes. These two modes may in turn correspond to different
critical temperatures and to figure out the actual mechanism of
chiral symmetry restoration
in the GVL the elaborated models including all these complications
are required.
\section* {Acknowledgments}

I would like to thank M. Harada for useful discussions.

\newpage
\bf {References}

\end{document}